\documentclass{imsart}

\usepackage{amsthm,amsmath,amssymb,epsfig,subfigure,graphics}
\usepackage{subfig}

\startlocaldefs

\def\Pr{{\rm I\!P}}
\def\be{\begin{equation}}
\def\ee{\end{equation}}
\def\bea{\begin{eqnarray*}}
\def\eea{\end{eqnarray*}}
\def\bean{\begin{eqnarray}}
\def\eean{\end{eqnarray}}

\def\ra{\rightarrow}

\def\Bl{\Bigl}
\def\Br{\Bigr}

\def\II{{\cal I}}

\def\eps{\epsilon}

\def\HC{\mbox{HC}_n^*}
\def\BJ{\mbox{BJ}_n^+}
\def\ALR{\mbox{ALR}_n}
\newtheorem*{Theorem}{Theorem}
\endlocaldefs

\begin{document}

\begin{frontmatter}

\title{The Average Likelihood Ratio for Large-scale Multiple
Testing and Detecting Sparse Mixtures}
\runtitle{Average Likelihood Ratio for Detecting Mixtures}

\author{\fnms{Guenther} \snm{Walther}\corref{}\ead[label=e1]{
gwalther@stanford.edu}\thanksref{t1}}
\thankstext{t1}{Work supported by NSF grant DMS-1007722} 
\address{390 Serra Mall\\ Stanford, CA 94305\\ \printead{e1}}
\affiliation{Stanford University}

\runauthor{G. Walther}

\begin{abstract}
Large-scale multiple testing problems require the simultaneous
assessment of many p-values. This paper compares several methods
to assess the evidence in multiple binomial counts of p-values:
the maximum of the binomial counts after standardization (the
`higher-criticism statistic'), the maximum of the binomial counts
after a log-likelihood ratio transformation (the `Berk-Jones statistic'),
and a newly introduced average of the binomial counts after a
likelihood ratio transformation. Simulations show that the
higher criticism statistic has a superior performance to the
Berk-Jones statistic in the case of very sparse alternatives
(sparsity coefficient $\beta \gtrapprox 0.75$), while the situation
is reversed for $\beta \lessapprox 0.75$. The average likelihood
ratio is found to combine the favorable performance of higher
criticism in the very sparse case with that of the Berk-Jones
statistic in the less sparse case and thus appears to dominate
both statistics. Some
asymptotic optimality theory is considered but found to set in
too slowly to illuminate the above findings, at least for sample
sizes up to one million. In contrast, asymptotic approximations
to the critical values of the Berk-Jones statistic that have
been developed by Wellner and Koltchinskii~(2003) and Jager and
Wellner~(2007) are found to give surprisingly accurate approximations
even for quite small sample sizes.
\end{abstract}



\end{frontmatter}

\section{Introduction}

This paper is concerned with the following mixture problem: One 
observes $X_1,\ldots,X_n$ i.i.d. $F$ and one wants to test
\begin{align*}
H_0:\ &F=\Phi,\ \ \ \mbox{ the standard normal distribution function}\\
 &\mbox{versus}\\
H_1:\ &F=(1-\eps) \Phi + \eps \Phi( \cdot -\mu)\ \ \ \ \mbox{ for some }
\eps \in (0,1), \mu>0.
\end{align*}

Interest in this prototypical setting derives from a number
of applications that involve large-scale multiple testing,
see e.g. Donoho and Jin~(2004). In the case where
the proportion of nonzero means is small, $\eps=\eps_n=n^{-\beta}$,
for $\beta \in (\frac{1}{2},1)$, there is
the following result: Parametrize $\mu=\mu_n=\sqrt{2r \log n}$
for $r \in (0,1)$ and define the detection boundary
\begin{equation*}
\rho^*(\beta)=
\begin{cases}
  \beta -\frac{1}{2} & \text{if $\frac{1}{2} < \beta \leq \frac{3}{4}$},\\
  (1-\sqrt{1-\beta})^2 & \text{if $\frac{3}{4} < \beta < 1$}.
\end{cases}
\end{equation*}

If $r<\rho^*(\beta)$, then it is impossible to detect the presence
of the nonzero means $\mu_n$: Any test with asymptotic level $\alpha \in (0,1)$
can only have trivial asymptotic power $\alpha$. On the other hand,
 if $r>\rho^*(\beta)$,
then the likelihood ratio test (which requires the knowledge of $\beta$
and $r$) at asymptotic level $\alpha$ will have asymptotic power 1,
see Ingster~(1997,1998) and Jin~(2004). But $\beta$ and $r$ are unknown,
so direct application the likelihood ratio test is not possible.
Jin~(2004) and Donoho and Jin~(2004) propose to employ the higher criticism
statistic
$$
\HC\ = \ \max_{1 \leq i \leq n/2} \sqrt{n} \Bigl(i/n -p_{(i)}\Bigr)
/\sqrt{p_{(i)}(1-p_{(i)})},
$$
where $p_i=\Pr(N(0,1)>X_i)$ is the p-value of $X_i$, and they show
that $\HC$ also attains the optimal detection boundary, i.e. $\HC$
has asymptotic power 1 for all $\beta \in (\frac{1}{2},1)$ and 
$r>\rho^*(\beta)$. Note that $\HC$
does not require the knowledge of $\beta$ and $r$.

\section{Combining the evidence of multiple binomial counts}
\label{binomial}

Denote by $F_n$ the empirical distribution function of the p-values:
$F_n(t):=\frac{1}{n} \sum_{i=1}^n 1(p_i \leq t)$. Then one sees that
\be  \label{HC}
\HC = \max_{t\in \{p_{(1)},\ldots, p_{(n/2)}\}} \sqrt{n} \frac{F_n(t)-t}{
\sqrt{t(1-t)}}.
\ee
Under the null hypothesis, the p-values $p_i$ are an i.i.d. sample
from $U[0,1]$. Thus the quantity $\sqrt{n} \frac{F_n(t)-t}{
\sqrt{t(1-t)}}$ is the standardized count of p-values that fall
in the interval $(0,t]$, and so $\HC$ looks for an excessive number
of p-values in the intervals $(0,t]$ for $t\in (0,\frac{1}{2}]$
by considering the maximum of these
standardized binomial counts over the intervals $(0,p_{(i)}]$ for
$i=1,\ldots,n/2$. 

While a standardized
binomial random variable is a classical example to illustrate
the convergence to a normal distribution, it is important to keep
in mind that its long tail is not any more subgaussian:
As the success probability moves from $\frac{1}{2}$ to $0$, 
the long tail becomes increasingly heavy,
see Shorack and Wellner~(1986,Ch.11.1). In fact, the first several
terms in $\HC$ even have heavy algebraic tails, as can be seen from
an argument similar to Sec.~3 in Donoho and Jin~(2004). Since the
distribution of the $\max$ depends sensitively on the tails, this
means that standardizing the counts does not guarantee that all counts
are treated equally. Rather, $\HC$
gives increasingly more weight
to counts with smaller index $i$. This raises the question what
effect this has on the performance of $\HC$.

To investigate this issue, we can compare the performance
of $\HC$ with a statistic that standardizes the binomial counts
differently to avoid unequal and heavy tails. Such a standardization
is given by the log-likelihood ratio transformation. Define
\begin{equation*}
logLR_n(t)=
\begin{cases}
  nF_n(t) \log \frac{F_n(t)}{t} + n(1-F_n(t))\log \frac{1-F_n(t)}{1-t}
    & \text{if $0<t<F_n(t)$},\\
    0 & \text{otherwise}.
\end{cases}
\end{equation*}
$logLR_n(t)$ is the one-sided log-likelihood ratio statistic
for testing whether the parameter of the binomial count $nF_n(t)$
equals $t$ vs. whether it is larger than $t$. The log-likelihood
ratio transformation possesses
the important property that it produces clean subexponential tails under the
null hypothesis, no matter what the binomial parameter $t$. This fact is 
implicit in the proof of the Chernoff-Hoeffding theorem, see Hoeffding~(1963).
One can now proceed as with $\HC$ and take the maximum of the thus
standardized binomial counts over the random intervals $(0,p_{(i)}]$.
This essentially yields a statistic proposed by Berk and Jones~(1979):
$$
\BJ\ =\ \max_{1\leq i \leq n/2} logLR_{n,i},
$$
where $logLR_{n,i}:=logLR_n(p_{(i)})=(i \log \frac{i}{np_{(i)}} +
(n-i) \log \frac{1-i/n}{1-p_{(i)}}) 1(p_{(i)}<\frac{i}{n})$. 
$\BJ$ was shown by Donoho
and Jin~(2004) to also attain the optimal detection boundary.
Both $\HC$ and $\BJ$ are
special cases of a family of goodness-of-fit tests based on 
$\phi$-divergences that are
introduced and studied by Jager and Wellner~(2007).

\begin{figure}[h]
\centering
\subfigure{
 \includegraphics[width=0.45\textwidth,height=0.28\textheight,trim = 22mm 70mm 24mm 75mm,clip]{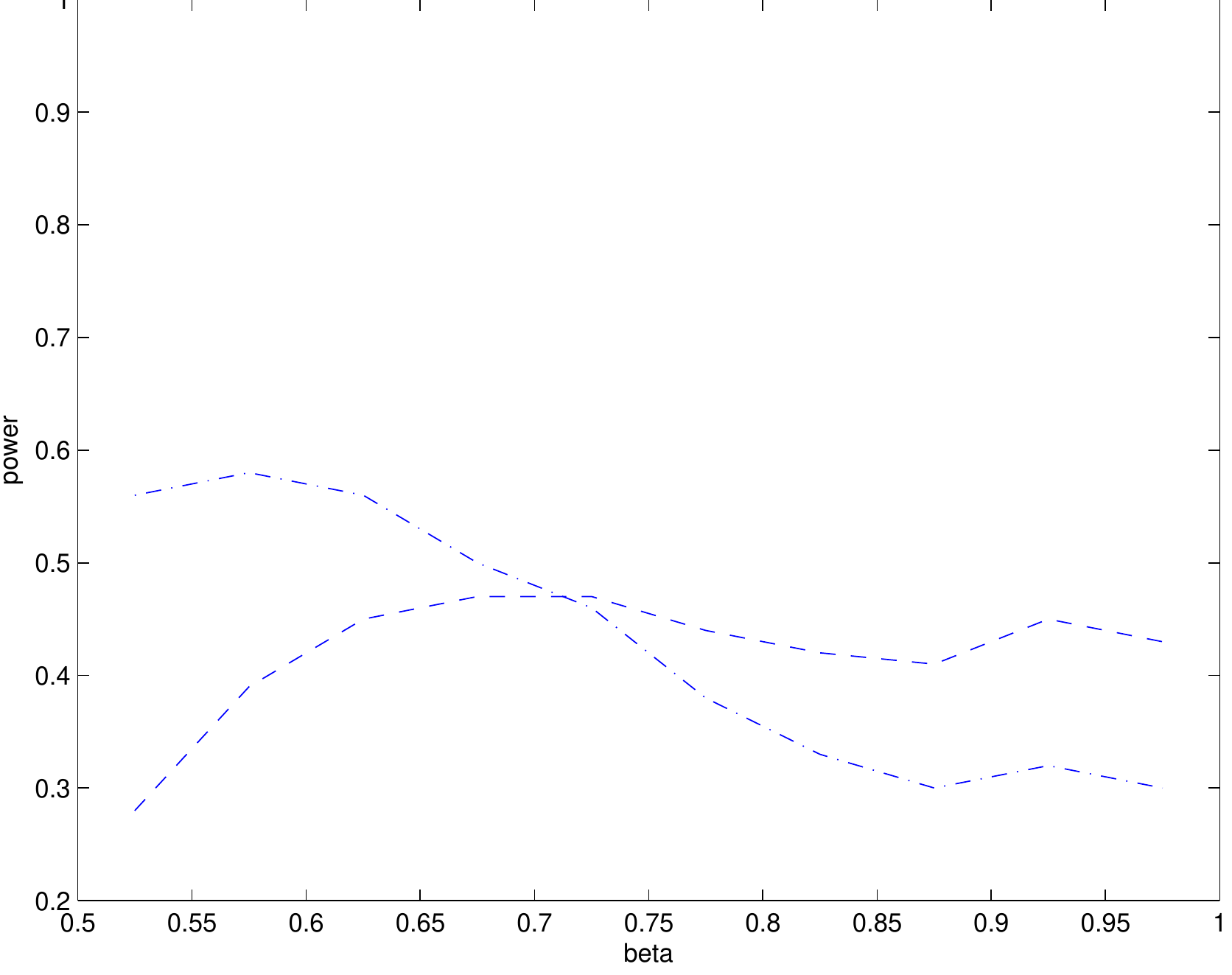}}
\subfigure{
 \includegraphics[width=0.45\textwidth,height=0.28\textheight,trim = 22mm 70mm 24mm 75mm,clip]{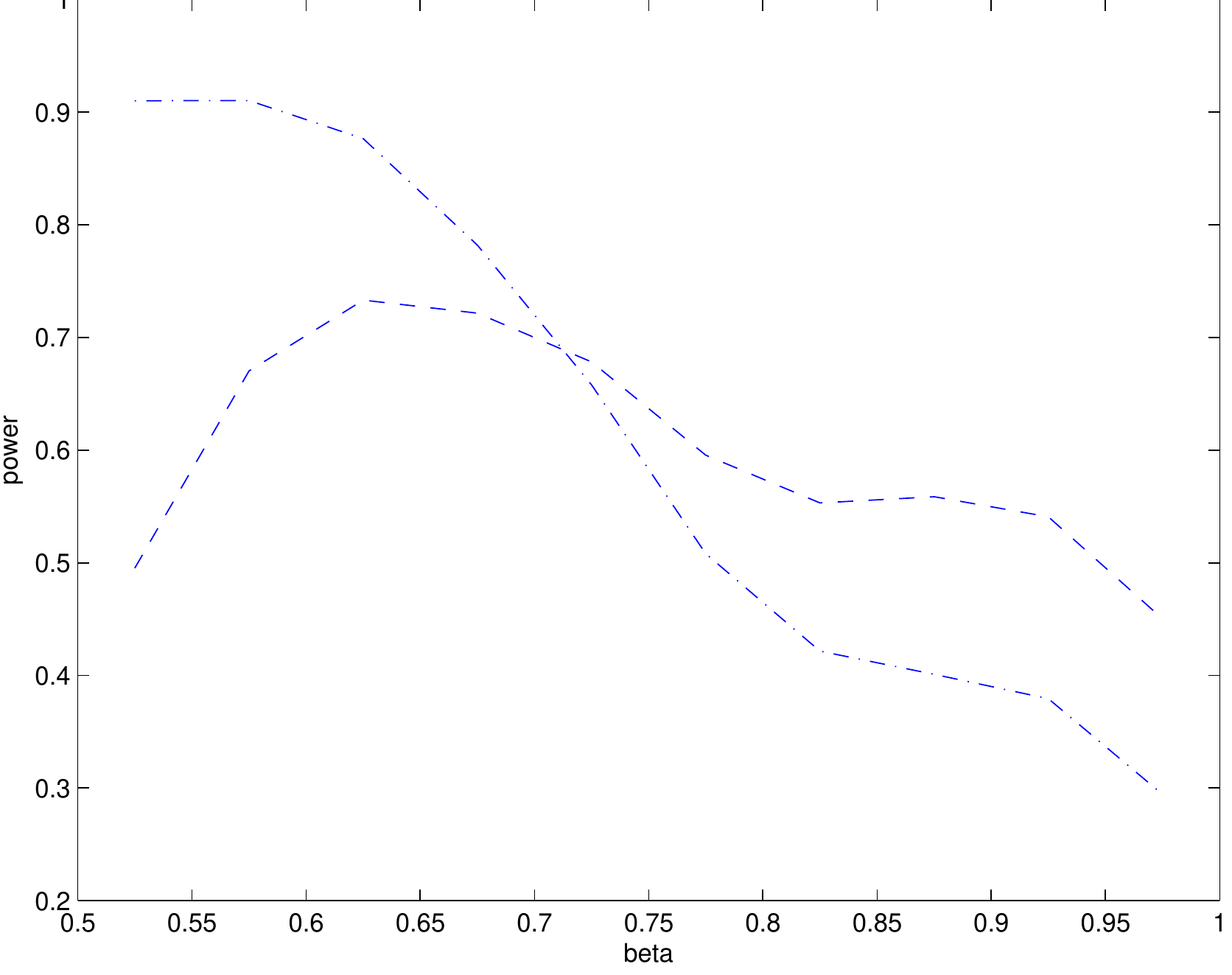}}\\
\caption{Power of $\HC$ (dashed) and $\BJ$ (dash-dot) as a function
of the sparsity parameter $\beta$. The left plot shows power for sample size
$n=10^4$, the right plot for $n=10^6$.}
\label{fig1}
\end{figure}

We compare the power of $\HC$ and $\BJ$ against alternatives
$\mu_n=\sqrt{2r \log n}$ with $r=r(\beta)=1.2 \rho^*(\beta)+0.1$ for
ten equally spaced values of $\beta$ between 0.5 and 1. The significance
level was set to 5\% by estimating the exact finite sample
critical values of $\HC$ and $\BJ$ with $10^5$ simulations. The power
of $\HC$ and $\BJ$ was then simulated with $10^4$ simulations.
The left plot in Figure~\ref{fig1} shows the resulting 
power values for sample size $n=10^4$,
the right plot for sample size $n=10^6$. One sees that $\HC$ has
a better detection performance in the very sparse case $\beta \gtrapprox 
\frac{3}{4}$,
while $\BJ$ has a better performance for smaller $\beta$.

The preceding discussion suggests the following explanation of this result:
Donoho and Jin~(2004) observed that for $\beta \in [\frac{3}{4},1)$
the strongest evidence against $H_0$ is found near the maximum of the
observations, i.e. at the smallest p-values. Since $\HC$ gives more
weight to smaller p-values compared to $\BJ$, $\HC$ will have more
power. But when $\beta \in (\frac{1}{2},\frac{3}{4})$, then the most 
informative place to look is at larger p-values, i.e. one needs to examine the 
count of p-values 
in the interval $(0,t]$ for certain $t \in (0,1)$. Since $\HC$ gives
less weight to the evidence in those intervals, it suffers a performance 
penalty in this case.

The simulation study also confirms the cautionary remarks in
Donoho and Jin~(2004) about the sample size required for the
above asymptotic optimality theory to adequately assess the
performance of statistical procedures. Both $\HC$ and $\BJ$ attain
the optimal detection boundary, i.e. have asymptotic power 1 against
the alternatives considered in the above simulation study. But even
for a sample size of one million, their detection power is quite
small for a large range of $\beta$ values. Moreover, the difference in
power between these two optimal procedures is larger than the
gain in power obtained by increasing the sample size 100fold from $n=10^4$
to $n=10^6$. Thus it appears that the asymptotic optimality theory
sets in too slowly to be informative for sample sizes up to at least a
million, and it seems prudent to instead assess the performance of
such procedures primarily via simulation studies.

The difference in performance between $\HC$ and $\BJ$ for various
$\beta$ raises the
question whether this difference represents an unavoidable trade-off, or 
whether it is possible to improve on this overall performance.
If a better performance is possible, how should one go about 
developing a better test?

\section{The average likelihood ratio statistic}

A promising approach to obtain good power uniformly in $\beta$
is a minimax test, which is typically constructed as a Bayes solution with
respect to a least favorable prior, see Lehmann and Romano~(2005,Ch.8.1).
But in the context at hand, such a construction appears to be involved
since it requires the specification a multivariate
prior over an appropriate set of alternative distributions. 

Instead we proceed as follows:
Suppose we start with an noninformative
uniform prior for the parameter $\beta$ on $(\frac{1}{2},1)$.
Given $\beta$, we can use knowledge about the problem to construct
an appropriate conditional test: 
Donoho and Jin~(2004) observe that for $\beta \in [\frac{3}{4},1)$
the most promising approach is essentially to look at the smallest p-value.
Thus we put prior probability $\frac{1}{2}$ on the likelihood ratio
test over the interval $(0,p_{(1)}]$.
For $\beta \in (\frac{1}{2},\frac{3}{4})$, the most promising interval to detect
alternatives with $r$ close to the detection boundary $\rho^*(\beta)=\beta-
\frac{1}{2}$
is the interval $(0,n^{-4r}]$. Thus given such a $\beta$, we will 
employ the likelihood ratio test on the interval $(0,t]$ with
$t=n^{-4(\beta-\frac{1}{2})}$. If $\beta \sim U(\frac{1}{2},\frac{3}{4})$,
then $t=n^{-4(\beta-\frac{1}{2})}$ has density proportional to $\frac{1}{t}$  
on $(\frac{1}{n},1)$.
Approximating the resulting posterior integral with the
corresponding weighted
sum of the $p_{(i)}$ and observing that the normalizing factor of
the weights is $\sum_{i=2}^{n/2} \frac{1}{i} \approx \log(n/3)$ yields
the {\sl average likelihood ratio}
$$
\ALR\ =\ \frac{1}{2} LR_{n,1} + \frac{1}{2} \sum_{i=2}^{n/2} 
\frac{1}{i \log(n/3)} LR_{n,i}
$$
where  
\begin{equation*}
LR_{n,i}=
\begin{cases}
    \Bigl( \frac{i}{n p_{(i)}}\Bigr)^i \Bigl(\frac{1- \frac{i}{n}}{1-p_{(i)}}
    \Bigr)^{n-i}
    & \text{if $p_{(i)}<\frac{i}{n}$},\\
    1 & \text{otherwise}.
\end{cases}
\end{equation*}
Thus $LR_{n,i}$ is the one-sided likelihood ratio statistic for testing
whether the parameter of the binomial count on $(0,t]$ equals $t$,
evaluated at $t=p_{(i)}$.

\begin{Theorem}
$\ALR$ attains the optimal detection boundary.
\end{Theorem}

For a proof, note that it was shown in Donoho and Jin~(2004) that
with probability converging to 1 there exists an index $i \in \{1,\ldots,
n/2\}$ such that $logLR_{n,i} \gtrapprox n^{\kappa}$, where $\kappa=
\kappa(\beta,r)>0$. Hence $\BJ$ (and $\HC$) grow algebraically fast 
under the alternative. Now $LR_{n,i}=\exp(logLR_{n,i})  \gtrapprox 
\exp(n^{\kappa})$. Thus $\ALR$ grows exponentially fast. Some informal
arguments given below suggest that $\ALR$ may have a limiting distribution 
under $H_0$, but to complete the proof in a rigorous way it is enough
to employ the upper bound $\ALR \leq 2 \exp(\BJ)$ together with
$\BJ/\log \log n \stackrel{p}{\ra} 1$ under $H_0$, 
see Jager and Wellner~(2007,Thm.3.1). $\Box$

The exponential increase of $\ALR$ has to be taken with a grain of
salt. Depending on $\beta$ and $r$, the constant $\kappa(\beta,r)$ may be 
close to zero. Then an enormous $n$ is required for $LR_{n,i}$ to overcome the
divisor $i \log(n/3)$ if $i\geq 2$. Of course, the same calamity befalls
$\BJ$ and $\HC$, where the polynomial $n^{\kappa}$ needs to overcome
a critical value of order $\log \log n$.
This appears to be one of the reasons why the asymptotic
theory is so slow to take hold.

\begin{figure}[h]
\centering
\subfigure{
 \includegraphics[width=0.45\textwidth,height=0.28\textheight,trim = 22mm 70mm 24mm 75mm,clip]{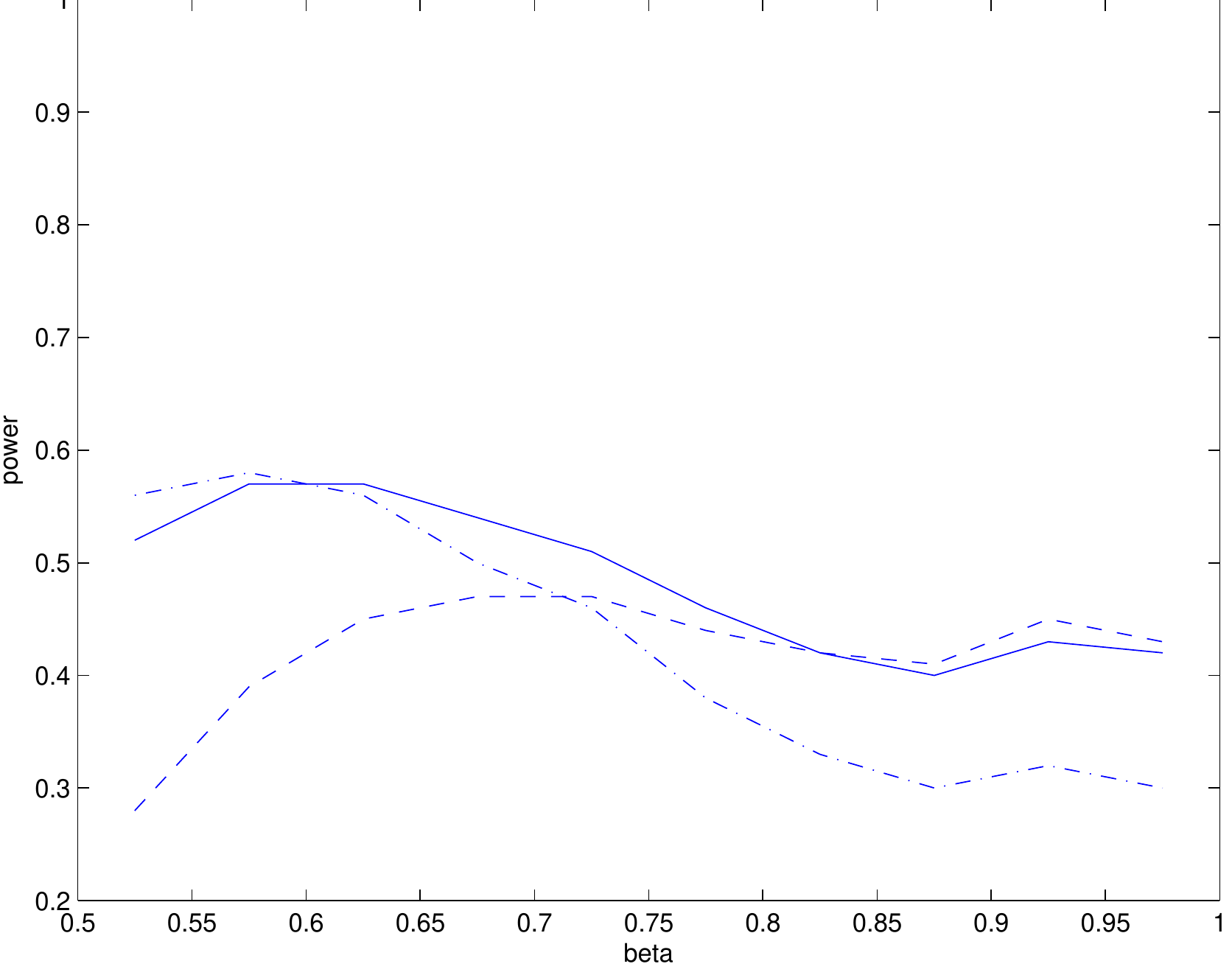}}
\subfigure{
 \includegraphics[width=0.45\textwidth,height=0.28\textheight,trim = 22mm 70mm 24mm 75mm,clip]{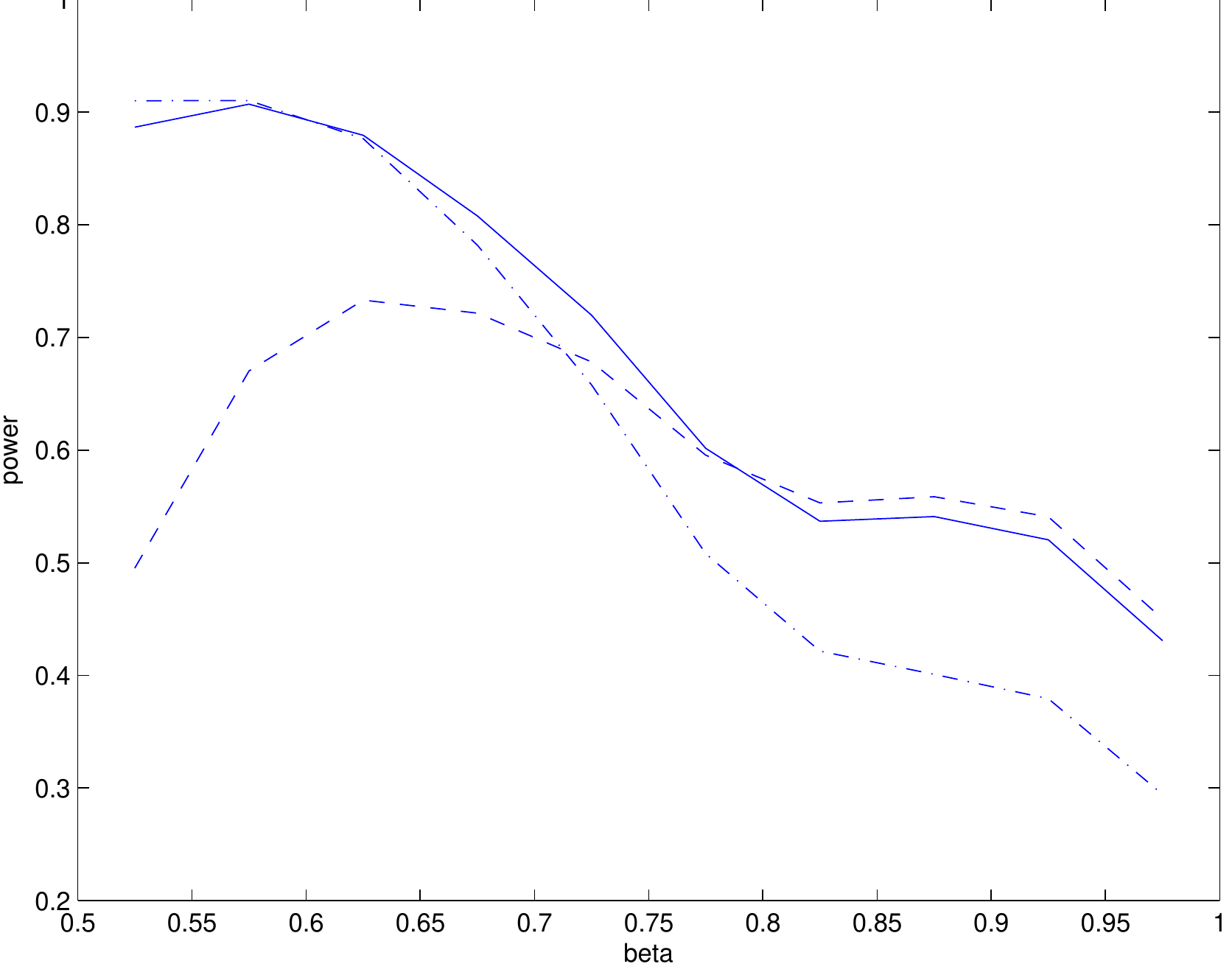}}\\
\caption{Power of $\ALR$ (solid), $\HC$ (dashed) and $\BJ$ (dash-dot) as a 
function of the sparsity parameter $\beta$. The left plot shows power for 
sample size $n=10^4$, the right plot for $n=10^6$.}
\label{fig2}
\end{figure}

As discussed above, it is therefore preferrable to evaluate the
performance of $\ALR$ with a simulation study. Figure~\ref{fig2}
compares the power of $\ALR$, $\HC$, and $\BJ$ in the same setting
that was considered in section~\ref{binomial}. 

One sees that $\ALR$
combines the good performance of $\HC$ at larger $\beta$ with the
good performance of $\BJ$ at smaller $\beta$ and thus results
in a test that appears to dominate both $\HC$ and $\BJ$.

To avoid numerical difficulties when $n$ is large, it is advisable
to rewrite $LR_{n,i}=\exp(logLR_{n,i})$ with $logLR_{n,i}$ given in
section~\ref{binomial}. As above, the simulation study
used a size of 5\% for all three tests by estimating the exact 
finite sample critical values with $10^5$ simulations. Since such a 
simulation may not be practical for larger samples, it is of interest
to explore whether reasonably accurate asymptotic approximations
are available.

\section{Asymptotic approximations for the null distributions}

A first attempt to derive a simple large sample approximation for the 
critical values of $\HC$ and
$\BJ$ can be  based on
$\HC/\sqrt{2 \log \log n} \stackrel{P}{\rightarrow} 1$
and $\BJ/\log \log n \stackrel{P}{\rightarrow} 1$, which follows e.g. from
Jager and Wellner~(2007,Thm.3.1). The significance levels obtained by
using the resulting thresholds $\sqrt{2 \log \log n}$
and $\log \log n$ for $\HC$ and $\BJ$, respectively, are listed under
`thresh' in Table~\ref{table1}. One sees 
that the resulting size of the tests is very large even for $n=10^6$.

A more refined approximation can be derived from results about the convergence
to an extreme value distribution. In the case of $\HC$, this result
follows from Jaeschke~(1979) and Eicker~(1979), see also Shorack and 
Wellner~(1986, Ch.16). In the case of $\BJ$
a proof was sketched in Berk and Jones~(1979). Wellner and Koltchinskii~(2003)
note an apparent error in that sketch and give a rigorous proof. See also
Jager and Wellner~(2007,Thm.3.1) for a unified treatment of $\HC$ and $BJ$.
The latter theorem establishes convergence of two-sided versions
of $\BJ$ and $\frac{1}{2}(\HC)^2$, after centering,
 to an extreme value distribution with distribution function $E_v^4(x)=
\exp(-4\exp(-x))$. As remarked in Shorack and Wellner~(1986,p.600), the
two one sided versions as well as the two halves ($i \lessgtr n/2$)
are asymptotically independent. Therefore the pertinent limit for
$\HC$ and $\BJ$ considered here should be $E_v^1$. The resulting
approximation for the level $\alpha$ critical value for $\BJ$ is
\be \label{q}
q_{\alpha}:= \log \log n +\frac{1}{2} \log \log \log n -
\frac{1}{2} \log(4\pi)-\log(-\log(1-\alpha)),
\ee
and the corresponding approximation for $\HC$ is $\sqrt{2 q_{\alpha}}$.
It is known that convergence to an extreme value distribution is typically
extremely slow, see Hall~(1979). Thus there would seem to be little hope
that the above approximation is useful for moderate sample sizes, in
particular since it involves a doubly-iterated (!) logarithm.
But surprisingly, the simulation study in Table~\ref{table1}
shows that the above approximation (labelled `EVI') is quite good for
$\BJ$ even for sample sizes as small as $n=100$. This appears to be another
benefit of the clean exponential tails resulting from the
log-likelihood ratio transformation. Unfortunately, the
approximation does not work well for $\HC$, where it yields
very anti-conservative results.

Wellner and Koltchinskii~(2003) suggest a further improvement for 
the approximation to $\BJ$ by using
the centering $c_n^2/(2b_n^2)$ with $c_n=2 \log \log n 
+\frac{1}{2} \log \log \log n -\frac{1}{2} \log(4\pi)$ and 
$b_n^2=2 \log \log n$ in place of the first three terms on the right hand
side of (\ref{q}). The results of this approximation are labelled `EVII' in 
Table~\ref{table1}
and show a further improvement for $\BJ$, but still not a useful
outcome for $\HC$. This is presumably due to the heavy binomial
tails which are not taken care of by the standardization in $\HC$.

In connection to this it is worth pointing out that a key argument in
proving the above limit theorems is to show that with high probability
the first $\log^5 n$ terms in $\HC$ and $\BJ$ do not contribute to
the maximum, and that for the remaining terms a strong approximation
with a Brownian bridge is applicable.
In particular, this means
that asymptotically the heavy binomial tails don't matter, and that the
maximum will not be attained at the first few terms. But as shown
by the simulations above and elsewhere, such as in Donoho and Jin~(2004),
this is certainly not the case for sample sizes of up to at least $n=10^6$,
which is the largest sample size we could explore in a reasonable amount
of time.
As remarked in Wellner~(2006,p.43) concerning the applicability of the
asymptotic results,
one needs $n > 1010388 \approx 10^6$
just to get $\log^5 n < n/2$.

\begin{table} 
\begin{tabular}{r|r|r|r|r|r|r|r|r|r|r}
Calibration & \multicolumn{2}{c|}{thresh} & \multicolumn{4}{c|}{EVI} &
\multicolumn{4}{c}{EVII} \\ \hline
Statistic & $\HC$ & $\BJ$ &
\multicolumn{2}{c|}{$\HC$} & \multicolumn{2}{c|}{$\BJ$} &
\multicolumn{2}{c|}{$\HC$} & \multicolumn{2}{c}{$\BJ$} \\
Nominal level in \%&- &- & 5 & 10 & 5 & 10 &
5 & 10 & 5 & 10 \\ \hline
$n=10^2$ & 44.7 & 34.7 & 20.8 & 27.2 & 7.2 & 13.4 & 19.6 & 25.3 & 6.2 & 11.4\\
$10^3$  & 45.0 & 34.0 & 20.0 & 26.2 & 6.7 & 12.3 & 19.1 & 25.1 & 6.1 & 11.2\\
$10^4$  & 45.7 & 34.4 & 19.2 & 25.2 & 6.4 & 11.7 & 18.6 & 24.3 & 5.9 & 10.9\\
$10^5$  & 45.6 & 34.4 & 18.4 & 24.4 & 6.2 & 11.3 & 17.9 & 23.7 & 5.9 & 10.7\\
$10^6$  & 46.0 & 34.9 & 18.0 & 23.9 & 6.2 & 11.4 & 17.6 & 23.3 & 5.9 & 10.8
\end{tabular}
\caption{Finite sample significance levels (in \%) of $\HC$ and $\BJ$
for various asymptotic approximations to critical values. Based on $10^5$
simulations.}
\label{table1}
\end{table}

Next we consider $\ALR$ and write
\bea
logLR_{n,1} & = & \Bl[\log \frac{1}{np_{(1)}} +(n-1)\log \frac{1-1/n}{
1-p_{(1)}} \Bigr] 1(p_{(1)} <1/n)\\
& = & \Bl[-\log\Bigl(np_{(1)} \Bl(1-\frac{np_{(1)}}{n} \Br)^n\Br)
  + \log (1-p_{(1)})+(n-1) \log(1-1/n)\Br] 1(np_{(1)}<1).
\eea
Recall that under $H_0$ we can use the
representation $p_{(1)}\stackrel{d}{=} E_1/(E_1+\ldots +E_{n+1})$,
where $\{E_i\}$
is an infinite sequence of i.i.d. Exp(1) random variables, see
Shorack and Wellner~(1986,p.335). Thus $logLR_{n,1}$ has the same
distribution as a random variable that converges a.s. to
$(-\log E_1 +E_1 -1)1(E_1<1)$ by the strong law.
Hence
\be  \label{LR1}
LR_{n,1} \stackrel{d}{\ra} \Bl(\frac{\exp(E_1)}{eE_1}\Br)^{
1(E_1<1)}.
\ee

Next, set $\II_n:= \{i: p_{(i)} \leq \log^5n/n\}$. Using (A.4) in
Donoho and Jin~(2004) and (26) on p.602 of Shorack and Wellner~(1986),
we get
$$
\max_{i\in \II_n} \ logLR_{n,i} \ \leq \ \max_{i\in \II_n} 
\frac{\Bl(\frac{i}{n} -p_{(i)}\Br)^2}{2p_{(i)}(1-p_{(i)})}\ =
\ o_p(\log \log n).
$$
Hence on the event ${\cal A}_n:=\{ \# \II_n \leq 2 \log^5 n\}$:
$$
\sum_{i \in \II_n} \frac{1}{i \log(n/3)} LR_i \ \leq \ \exp\Bl(
(o_p(\log \log n)\Br) \frac{2\log \log n}{\log(n/3)}
\ =\ o_p(1),
$$
and $\Pr({\cal A}_n^c)=\Pr(\mbox{bin}(n,\log^5n/n)>2\log^5n) \ra 0$
by Chebychev.

For $p_{(i)}>\log^5n/n$ one can proceed as in the proof of Thm.~3.1
in Jager and Wellner~(2007), see also the proof of Thm.~1.1 in
Wellner and Koltchinskii~(2003), and as on p.601 of Shorack and Wellner~(1986)
and first approximate the log-likelihood ratio process by the square of the
normalized empirical process and then by the square of a normalized
Brownian Bridge. This suggests that
$$
\sum_{i=2}^{n/2} \frac{1}{i \log(n/3)} LR_{n,1} \approx L_n:=
\frac{1}{\log n} \int_{1/n}^{1/2} \frac{1}{t} \exp\Bl(
\frac{{B^+}^2(t)}{2t(1-t)}\Br) dt.
$$
It is not clear whether $L_n$ has a finite limit distribution.
Simulations show that the quantiles of $L_n$ increase very slowly
as $n$ increases from $10^2$ to $10^6$.
Formally applying l'H\^{o}pital's rule gives $\lim_{n \ra \infty} L_n
=\lim_{n\ra \infty} \exp\Bl(\frac{{B^+}^2(1/n)}{2/n(1-1/n)}\Br)$.
Since $\exp\Bl( \frac{{B^+}^2(1/n)}{2/n(1-1/n)}\Br)
\stackrel{d}{=} \exp(\frac{1}{2}{Z^+}^2)$ with $Z \sim $N(0,1), a
conjecture for the limit law of $\ALR$ would be

\be  \label{ALRlimit}
\frac{1}{2} \Bl(\frac{\exp(E_1)}{eE_1}\Br)^{1(E_1<1)} +
\frac{1}{2} \exp(\frac{1}{2}{Z^+}^2).
\ee

This expression reflects the fact that the beta distribution of the
first order statistic behaves like an exponential distribution, while
sufficiently larger order statistics possess a beta distribution
that is closer to a normal.
Of course, l'H\^{o}pital's rule is not applicable since
$\lim_{n\ra \infty} \exp\Bl( \frac{(B^+)^2(1/n)}{2/n(1-1/n)}\Br)$
does not exist by the law of the iterated logarithm for the Brownian
bridge, so even if the law of $L_n$
converges, the limit does not have to be the law of
$\exp(\frac{1}{2}{Z^+}^2)$.

Table~\ref{table2} gives the finite sample significance levels
of $\ALR$ resulting from the approximation (\ref{ALRlimit}) in 
the column `Calibration~1'.
The critical values used for calibration~1 are 6.05 and 3.42, which were 
obtained from $10^5$ simulations of (\ref{ALRlimit}).
Calibration~2 uses $L_n$ with $n=10^5$ in place of 
$\exp(\frac{1}{2}{Z^+}^2)$. The resulting critical values are
6.16 and 3.60. One sees that both approximations
are reasonably accurate, albeit somewhat anti-conservative,
 for the sample sizes considered.

\begin{table} 
\begin{tabular}{r|r|r|r|r}
 & \multicolumn{2}{c|}{Calibration 1} & \multicolumn{2}{c}{Calibration 2}\\ 
\hline
Nominal level in \%& 5 & 10 & 5 & 10 \\ \hline
$n=10^2$ & 6.3 & 12.5 & 6.2 & 11.7\\      
$10^3$ &  6.0  & 12.0 & 5.9 & 11.3\\      
$10^4$ &  5.8  & 11.9 & 5.7 & 11.1\\      
$10^5$ &  5.7  & 11.7 & 5.6 & 11.0\\      
$10^6$ &  5.7  & 11.8 & 5.4 & 11.0  
\end{tabular}
\caption{Finite sample significance levels (in \%) of $\ALR$
for two different approximations to the critical values of $\ALR$. 
Based on $10^5$
simulations.}
\label{table2}
\end{table}

\section{Relation to other work and open problems}

Different variations of the average likelihood ratio
have been
used successfully in other detection problems, see e.g. Shiryaev~(1963),
Burnashev and Begmatov~(1990), D\"{u}mbgen~(1998),
Siegmund~(2001), Gangnon and Clayton~(2001), Chan~(2009)
or Chan and Walther~(2011), but the above weighted average likelihood
ratio seems not to have been considered before.

It is worthwhile to compare the above results with the setting where the
proportion $\eps_n$ of observations with nonzero means is not scattered
randomly but possesses structure, e.g. when
$\eps_n n$ consecutive observations possess an elevated mean. Such
problems are typically addressed with the scan statistic, i.e.
the maximum likelihood ratio statistic. It was shown by Arias-Castro
et al.~(2005) that the scan can detect elevated means of size 
$\mu_n=\sqrt{2\log n/(\eps_n n)}$. Chan and Walther~(2011) showed
that the scan cannot do better than that but that a version of the
average likelihood ratio can detect smaller means where the factor
$\sqrt{2\log n}$ in the numerator 
is replaced by $\sqrt{2\log(1/\eps_n )}=\sqrt{2\beta\log n}$. 
No test can improve on this latter rate. Thus the scan is
optimal only in the case of a single elevated mean, but its performance
relative to the ALR deteriorates as the proportion of nonzero means
increases. It was also shown in Walther~(2010) and Chan and Walther~(2011) 
that optimality
of the scan can be restored by employing scale-dependent critical values.
Comparing with the results in the present paper, one sees that structure
in the elevated means allows to greatly improve the detection power:
In the case of consecutively elevated means, the detection boundary
is lowered by a factor $\sim \sqrt{\eps_n n}=\sqrt{n^{1-\beta}}$, which can be 
considerable.

Regarding the setting in the present paper, it would be of interest
to develop an optimality theory that allows to compare
the performance of tests at more moderate sample sizes. Such a comparison
might by possible by exploring the rate at which an estimator can
approach the detection boundary while still guaranteeing consistency.
See Walther~(2010) and Chan and Walther~(2011) for such an analysis
in the case of consecutively elevated means. Finally, it would be
of interest to perform a more formal 
investigation of a possible limit distribution of the average likelihood
ratio.

\subsection*{Acknowledgement} The author would like to thank
David Siegmund and Jon Wellner for helpful discussions.

\subsection*{References}

\begin{description}
\item[]
\textsc{Arias-Castro, E., Donoho, D.L.} and \textsc{Huo, X.} (2005). 
Near-optimal detection of geometric objects by fast multiscale methods.
\textit{IEEE Trans. Inform. Th.}
\textbf{51} 2402--2425.

\item[]
\textsc{Berk, R.H.} and \textsc{Jones, D.H.} (1979).
Goodness-of-fit test statistics that dominate the Kolmogorov
statistics. 
\textit{Z. Wahrsch. Verw. Gebiete}.
\textbf{47} 47--59.

\item[]
\textsc{Burnashev, M.V.} and \textsc{Begmatov, I.A.} (1990). 
On a problem of detecting a signal that leads
to stable distributions. 
\textit{Theory Probab. Appl.}
\textbf{35} 556--560.

\item[]
\textsc{Chan, H.P.} (2009). 
Detection of spatial clustering with average likelihood ratio test statistics.
\textit{Ann. Statist.}
\textbf{37} 3985--4010.

\item[]
\textsc{Chan, H.P.} and \textsc{Walther,G.} (2011).
Detection with the scan and the average likelihood ratio.
Manuscript.

\item[]
\textsc{Donoho, D.} and \textsc{Jin, J.} (2004).
Higher criticism for detecting sparse heterogeneous mixtures.
\textit{Ann. Statist.}
\textbf{32} 962--994.

\item[]
\textsc{D\"{u}mbgen, L.} (1998). 
New goodness-of-fit tests and their application to nonparametric confidence
sets. 
\textit{Ann. Statist.}
\textbf{26} 288--314.

\item[]
\textsc{Eicker, F.} (1979).
The asymptotic distribution of the suprema of the standardized
empirical processes.
\textit{Ann. Statist.}
\textbf{7} 116--138.

\item[]
\textsc{Gangnon, R.E. and Clayton, M.K.} (2001).
The weighted average likelihood ratio test for spatial disease
clustering. 
\textit{Statistics in Medicine}
\textbf{20} 2977--2987.

\item[]
\textsc{Hall, P.} (1979).
On the rate of convergence of normal extremes. 
\textit{J. Appl. Probab.}
\textbf{16}, 433--439.

\item[]
\textsc{Hoeffding, W.} (1963).
Probability inequalities for sums of bounded random variables.
\textit{J. Amer. Statist. Assoc.}
\textbf{58} 13--30.

\item[]
\textsc{Ingster, Y. I.} (1997). 
Some problems of hypothesis testing leading to infinitely divisible
distributions.
\textit{Math. Methods Statist.}
\textbf{6} 47--69.

\item[]
\textsc{Ingster, Y. I.} (1998).
Minimax detection of a signal for $l^n$-balls. 
\textit{Math. Methods Statist.}
\textbf{7} 401--428.

\item[]
\textsc{Jaeschke, D.} (1979).
The asymptotic distribution of the supremum of the standardized empirical
distribution function on subintervals.
\textit{Ann. Statist.}
\textbf{7} 108--115.

\item[]
\textsc{Jager, L.} and \textsc{Wellner, J.A.} (2007).
Goodness-of-fit tests via phi-divergences.
\textit{Ann. Statist.}
\textbf{35} 2018--2053.

\item[]
\textsc{Jin, J.} (2004).
Detecting a target in very noisy data from multiple looks.
\textit{IMS Monograph}.
\textbf{45} 255--286.

\item[]
\textsc{Lehmann, E.L.} and \textsc{Romano, J.P.} (2005).
\textit{Testing Statistical Hypotheses}, Third Edition, Springer, New York.

\item[]
\textsc{Shiryaev, A.N.} (1963)
On optimum methods in quickest detection problems.
\textit{Theory Probab. Appl.}
\textbf{8} 22-46.

\item[]
\textsc{Shorack, G.R.} and \textsc{Wellner, J.A.} (1986).
\textit{Empirical Processes with Applications to Statistics}.
Wiley, New York.

\item[]
\textsc{Siegmund, D.} (2001). 
Is peak height sufficient?
\textit{Genetic Epidemiology}
\textbf{20} 403--408.

\item[]
\textsc{Walther, G.} (2010).
Optimal and fast detection of spatial clusters with scan statistics. 
\textit{Ann. Statist.}
\textbf{38} 1010-1033.

\item[]
\textsc{Wellner, J.A.} (2006)
Goodness of fit via phi-divergences:
a new family of test statistics. Talk at Northwest Probability Seminar.
University of Washington, Seattle. October 22, 2006.

\item[]
\textsc{Wellner, J.A.} and \textsc{Koltchinskii, V.} (2003)
A note on the asymptotic distribution of
Berk-Jones type statistics under the null hypothesis. 
In \textit{High Dimensional Probability III}
(J. Hoffmann-Jorgensen, M. B. Marcus and J. A. Wellner, eds.) 321-332. 
Birkh\"{a}user, Basel.
\end{description}
\end{document}